\documentclass[preprint,showpacs,showkeys,preprintnumbers,amsmath,amssymb,superscriptaddress]{revtex4}

\usepackage{graphicx}
\usepackage{dcolumn}
\usepackage{bm}

\begin{document}

\author{R. Rossi Jr.}
\affiliation{Universidade Federal de S\~{a}o Jo\~{a}o del-Rei,
Campus Alto Paraopeba, C.P. 131, 36420–000, Ouro Branco, MG, Brazil}
\author{K.M. Fonseca Romero}
\affiliation{Departamento de F´ýsica, Facultad de Ciencias,
Universidad Nacional, Ciudad Universitaria, Bogot´a, Colombia}
\author{M. C. Nemes}
\affiliation{Departamento de F\'{\i}sica, Instituto de Ci\^{e}ncias
Exatas, Universidade Federal de Minas Gerais, C.P. 702, 30161-970,
Belo Horizonte, MG, Brazil}

\title{Semiclassical Dynamics from Zeno-Like measurements}

\begin{abstract}
The usual semiclassical approximation for atom-field dynamics
consists in substituting the field operators by complex numbers
related to the (supposedly large enough) intensity of the field. We
show that a semiclassical evolution for coupled systems can always
be obtained by frequent Zeno-like measurements on the state of one
subsystems, independently of the field intensity in the example
given. We study the Jaynes–Cummings model from this perspective.
\end{abstract}

\maketitle

The Quantum Zeno Effect was introduced in the literature by B. Misra
and E. C. Sudarshan as a paradox, a theoretical prediction, the
freezing of a quantum system by constant measurements, in
contradiction with experimental observations. The example presented
by the authors is the one of the traces of decaying particles in
bubble chambers \cite{art1}. In a later publication the same authors
solved the so call ``Quantum Zeno Paradox" \cite{art2}, arguing that
the observed tracks were not sufficiently frequent (could not be
considered as a continuous measurement), so it could not modify the
particle's lifetime.

In 1990, following Cook's proposal \cite{art3}, the first
experimental observation of the Quantum Zeno Effect was realized
\cite{art4}. The QZE became the center of interesting debates on
foundations of Quantum Mechanics \cite{art5,art6,art7}. This pioneer
experiment shows the inhibition of a quantum transition, differently
from the initial proposal of B. Misra and E. C. Sudarshan which
involved the freezing of a decaying dynamics. Recently an
experimental observation of the QZE on decaying systems was reported
\cite{art8,art9}.

The QZE is also studied in the context of quantum state protection
\cite{art10,art11,art12}, where an increasing number of protocols
and strategies proposed have shown the possibilities of practical
applications of QZE in quantum information processing.

Frequent measurements do not necessarily freeze the evolution of a
quantum system. They can induce an evolution restricted in subspaces
defined by the measurements. This evolution has been recently
investigated, in Ref. \cite{art13}, and called quantum Zeno
Dynamics. Strategies, based on quantum Zeno dynamics, to protect
quantum states from decoherence \cite{art14}, for state purification
\cite{art15} and distillation processes \cite{art16} have been
presented.

In the present contribution we show that the quantum Zeno dynamics
in a two degrees of freedom systems will inhibit entanglement and
render the dynamics in this sense semiclassical. We study the
dynamics of a two level atom interacting with an electromagnetic
mode, which is frozen in its initial state by frequent projective
measurements. The atomic subspace evolution becomes unitary in the
limit of infinite measurements, and depending on the measured field
state, the dynamics becomes identical to the semiclassical model for
this interaction. Therefore the quantum Zeno dynamics can justify
the semiclassical evolution.

Let us start by calculating the time evolution operator for the
quantum Zeno dynamics of a bipartite system, composed of subsystems
$A$ and $B$, where subsystem $B$ is frequently measured in its
initial state. The hamiltonian of the system is

\begin{equation}
H=H_{A}+H_{B}+H_{AB},
\end{equation}
where $H_{AB}$ is the interaction term. The initial global state is
given by:

\begin{equation}
\rho(0)=\rho_{A}(0)\otimes|B\rangle\langle B|,
\end{equation}
and its free time evolution (without measurements) can be written as

\begin{equation}
\rho(t)=e^{-\frac{i}{\hbar}[H,\bullet]t}\rho(0).
\end{equation}

The total dynamics is composed of free evolutions, through short
time intervals, followed by projective measurements in the state of
subsystem $B$. The total time of the evolution is $T=tN$ where $N$
is the number of measurements and $t$ is the free evolution time.
Notice that the time between measurements is inversely proportional
to the number of measurements. We consider a large number of
measurements, therefore the free evolution takes place in a very
short time interval, and write the series expansion in time for the
unitary time evolution up to second order terms $O(t^{2})$.

\begin{equation}
\rho(t)=\left(I-\frac{it}{\hbar}[H,\bullet]-\frac{t^{2}}{2\hbar^{2}}[H,[H,\bullet]]\right)\rho(0).
\end{equation}

The projective measurements in subsystem $B$ are represented by the
operator $P_{B}=I_{A}\otimes|B\rangle\langle B|$. After one
projective measurement $\rho(t)$ can be written as:

\begin{equation}
P_{B}\rho(t) P_{B}=\left\{I-\frac{it}{\hbar}\left[\langle
H\rangle_{B},\bullet\right]-\frac{t^{2}}{2\hbar^{2}}\left(\langle
H^{2}\rangle_{B}\bullet-\langle H \rangle_{B}\bullet \langle H
\rangle_{B}-\bullet\langle H^{2}
\rangle_{B}\right)\right\}\rho(0),\label{exp1}
\end{equation}
where $\langle H\rangle_{B}=\langle B| H |B \rangle$. The expression
in (\ref{exp1}) can also be written as:

\begin{equation}
\exp\left(-\frac{it}{\hbar}\left[\langle
H\rangle_{B},\bullet\right]-\frac{t^{2}}{2\hbar^{2}}\left(\langle
H^{2}\rangle_{B}\bullet-\langle H \rangle_{B}\bullet \langle H
\rangle_{B}-\bullet\langle H^{2}
\rangle_{B}\right)\right)\rho(0)=\left(e^{L_{A}t}\rho_{A}\right)\otimes|B\rangle\langle
B|,
\end{equation}
notice that $L_{A}=-\frac{it}{\hbar}\left[\langle
H\rangle_{B},\bullet\right]-\frac{t^{2}}{2\hbar^{2}}\left(\langle
H^{2}\rangle_{B}\bullet-\langle H \rangle_{B}\bullet \langle H
\rangle_{B}-\bullet\langle H^{2} \rangle_{B}\right)$ is a super
operator that acts only in subsystem $A$.

The state operator at time $T$ can be calculated after $N$
applications of $e^{L_{A}t}$ in $\rho(0)$:

\begin{equation}
\rho(T)=(e^{L_{A}t})^{N}\rho(0)
=\left(e^{NL_{A}t}\rho_{A}\right)\otimes|B\rangle\langle B|,
\end{equation}
as $t=T/N$, we can write

\begin{equation}
e^{NL_{A}t}=\exp\left(-\frac{iT}{\hbar}\left[\langle
H\rangle_{B},\bullet\right]-\frac{T^{2}}{N\hbar^{2}}\left(\langle
H^{2}\rangle_{B}\bullet-\langle H \rangle_{B}\bullet \langle H
\rangle_{B}-\bullet\langle H^{2} \rangle_{B}\right)\right)
\end{equation}
and in the limit $N\rightarrow\infty$

\begin{equation}
e^{NL_{A}t}=e^{L_{A}T}=\exp\left(-\frac{i}{\hbar}[\langle
H\rangle_{B},\bullet]T\right)\label{exp2}.
\end{equation}

The operator $e^{NL_{A}t}$ governs the quantum zeno dynamics on the
bipartite system. Notice that if subsystem $B$ is frozen, the
evolution of subsystem $A$ is unitary and depends on the projective
measurements in $B$. The coefficient of second order terms
$O(t^{2})$ tend to zero as $N\rightarrow\infty$, therefore, in this
limit there is no entanglement between the systems. Nevertheless,
the hamiltonian interaction term $H_{AB}$ may be significant for the
evolution of subsystem $A$. The expression (\ref{exp2}) confirm the
results presented in Ref. \cite{art15,art16}.

The measurements on subsystem $B$ are relevant for subsystem $A$
only when there is interaction between them ($H_{AB}\neq 0$). Next
we show that, for specific parameters, the quantum zeno dynamics may
describe the semiclassical interaction between a two level atom and
an electromagnetic mode.

The well known Jaynes-Cummings model describes the atom field
interaction in the quantum mechanical context. Subsystems that
represent the degrees of freedom of the atoms and the field have the
dynamics described by a hamiltonian which allows for excitation
transfer between the subsystems. In the semiclassical model for this
interaction only the atomic degrees of freedom are described as
quantum elements, the electromagnetic field is considered as an
external potential responsible for the atomic levels coupling.

We show how to obtain an evolution equivalent to the semiclassical
atom field dynamics from the quantum zeno dynamics on
Jaynes-Cummings model. Following the terminology presented before,
let us consider subsystem $A$ describing the atom and subsystem $B$
the field. We have already shown that quantum zeno dynamics prevent
entanglement between the subsystems, and restricts the evolution on
subsystem $A$. Both factors contributed to the construction of the
semiclassical interaction from quantum zeno dynamics.

It is well known that the Jaynes-Cummings model is equivalent to the
semiclassical model when the electromagnetic field is prepared in a
coherent state with a huge number of photons. The experimental
observation of this equivalence was reported in Ref.\cite{art17}.
This experiment consists in a atomic interferometer, where a field
mode, prepared in a coherent state and interacting with the atoms,
act as a ``beam-splitter". In the quantum regime (where the mean
photons number is low $|\alpha|^2\approx 1$) the field is able to
register the atomic ``path", the atom-field interaction produces an
entangled state. In classical regime (where the mean photon number
is high) the microscopic ``beam-splitter" does not record the
particle's path. The field state remains basically the same before
and after the interaction with the two level atom, the atomic energy
is not sufficient to modify the field state, this is an essencial
element that allows for a semiclassical dynamics between atom and
field.

In the quantum zeno dynamics of the Jaynes-Cummings model the field
state also remains in its initial state. Frequent projective
measurements inhibits the field evolution, therefore, as in the
semiclassical dynamics, the atom interacts with a system that does
not evolve, and can be seen as an external perturbation in the
atomic evolution.

The Jaynes-Cummings hamiltonian, in RWA, is given by

\begin{equation}
H_{JC}=\frac{1}{2}\hbar\omega_{a}\sigma_{z}+\hbar\omega
a^{\dagger}a+ g[a\sigma_{+}+a^{\dagger}\sigma_{-}]\label{hamcoh},
\end{equation}
where $\omega_{a}$ is the atomic transition frequency, $\omega$ is
the electromagnetic field frequency and $g$ is the coupling
coefficient.

As shown in equation (\ref{exp2}) the atomic subspace evolution
depends on the state $|B\rangle$ where the field is measured. If the
field is projected in the state $|B\rangle=|\alpha\rangle$, the
hamiltonian $\langle H_{JC}\rangle_{B}$ that governs the dynamics in
(\ref{exp2}) can be written as

\begin{equation}
\langle H_{JC}\rangle_{B} = \frac{1}{2}\hbar\omega_{a}\sigma_{z}+
g[\alpha\sigma_{+}+\alpha^{*}\sigma_{-}],\label{hamsemi2}
\end{equation}
that is the exact same form of the hamiltonian in semiclassical
description of mater and light interaction. Notice that this
``classical limit" does not depend on the coherent state mean photon
number, as long as the field state has its evolution inhibited by
projective measurements, the atomic dynamics is given by the
semiclassical Hamiltonian (\ref{hamsemi2}). The time evolution
operator of the quantum Zeno evolution, deduced in equation
(\ref{exp2}), governed by Hamiltonian (\ref{hamsemi2}) justify the
atom-field semiclassical dynamics.

If the field is projected in any linear combination of consecutive
fock states, as
$|B\rangle=\cos(\theta)|n\rangle+e^{i\phi}\sin(\theta)|n+1\rangle$,
the Hamiltonian responsible for the time evolution on the atomic
subsistem is given by

\begin{equation}
\langle H\rangle_{B}= \frac{1}{2}\hbar\omega_{a}\sigma_{z}+
g[\cos(\theta)e^{i\phi}\sin(\theta)\sqrt{n+1}\sigma_{+}+\cos(\theta)e^{-i\phi}\sin(\theta)\sqrt{n+1}\sigma_{-}].\label{mediahjc}
\end{equation}
The evolution governed by (\ref{mediahjc}) is similar to the one in
(\ref{hamsemi2}).

On the other hand, if the field is measured in a fock state, as
$|B\rangle=|n\rangle$, the hamiltonian $\langle H_{JC}\rangle_{B}$,
does not couple the atomic states any more.

\begin{equation}
\langle H_{JC}\rangle_{B}=\langle
n|(\frac{1}{2}\hbar\omega_{a}\sigma_{z}+\hbar\omega a^{\dagger}a+
g[a\sigma_{+}+a^{\dagger}\sigma_{-}])|n\rangle=\frac{1}{2}\hbar\omega_{a}\sigma_{z}+\hbar\omega.
\end{equation}
The reason is that the coupling term in $H_{JC}$ is linear,
therefore the average in Fock states on it is null.

The divergence between the evolution generated by projections of the
field in the states $|n\rangle$, $|\alpha\rangle$ and $
\cos(\theta)|n\rangle+e^{i\phi}\sin(\theta)|n+1\rangle$ are related
to the information obtained on subsistem $B$. Projections in
$|n\rangle$ represents complete information about the number of
photons in the electromagnetic mode. Therefore, the dynamics
composed by frequent measurements ($N\rightarrow\infty$) on
subsistem $B$ in $|n\rangle$ conserves this subsistem in a defined
photon number state. This dynamics does not allow for excitation
exchange between atoms and the field mode. On the other hand,
projections in $|\alpha\rangle$ or
$\cos(\theta)|n\rangle+e^{i\phi}\sin(\theta)|n+1\rangle$ represents
incomplete information about the photons number, and excitations
exchange are possible.

To summarize, we have shown that the quantum Zeno dynamics can
justify a semiclassical hamiltonian for the atom-field interaction
even in the small photon number limit. Projective measurements in
the field state induce an unitary evolution on atomic subspace. For
appropriate choices of the measured states this unitary evolution is
identical to the one governed by a semiclassical hamiltonian.

\end{document}